\documentclass[prb,twocolumn,groupedaddress,shownopacs,amssymb]{revtex4-2}
\usepackage{graphicx,latexsym}
\usepackage{amsmath,amssymb,amsfonts}
\usepackage{bm}
\usepackage{braket}
\usepackage{mathrsfs}
\usepackage{mathdots}

\newcommand{\up}{\uparrow}
\newcommand{\down}{\downarrow}
\newcommand{\m}{\mathcal}
\newcommand{\ep}{\varepsilon}
\begin{document}
\title{Effects of Nonmagnetic Impurities on Nonequilibrium Phase Transitions in Microwave-Driven Superconductors}

\author{Taira Kawamura} 
\affiliation{Department of Physics, College of Science and Technology, Nihon University, Tokyo 101-8308, Japan}

\date{\today}

\begin{abstract}
We investigate the nonequilibrium steady states of a microwave-driven BCS superconductor in the presence of nonmagnetic impurity scattering and energy relaxation. Using the Floquet--Nambu--Keldysh Green's function method together with the self-consistent Born approximation, we self-consistently determine the superconducting order parameter in the microwave-driven steady state. We find that the nonequilibrium phase diagram depends strongly on the impurity-scattering strength. When impurity scattering is absent or weak, the steady-state order parameter is single-valued at each temperature and drive frequency. At intermediate scattering strengths, we find bistability between microwave-enhanced and microwave-suppressed superconducting states. As the impurity-scattering strength increases further, we instead find bistability between the superconducting and normal states. This evolution reflects competition between two effects of microwave driving: Eliashberg-type enhancement through the redistribution of thermally excited quasiparticles and suppression through pair-breaking-induced quasiparticle generation. In contrast to the robustness of equilibrium $s$-wave superconductivity against nonmagnetic impurity scattering, our results demonstrate that nonmagnetic impurity scattering plays a central role in determining the character of nonequilibrium superconducting phase transitions under microwave driving.
\end{abstract}

\maketitle
\section{Introduction}
\label{sec.Intro}

The essential role of the quasiparticle distribution in nonequilibrium superconductivity was succinctly articulated by Schmid: ``the distribution of the quasiparticles plays a most important role in the phenomena of nonequilibrium superconductivity''~\cite{Schmid1981}. In thermal equilibrium, the quasiparticle distribution is ``the rather dull object known as the Fermi function''~\cite{Schmid1981}. Out of equilibrium, however, the quasiparticle distribution becomes an independent degree of freedom determined by the competition among energy absorption, inelastic relaxation, and elastic impurity scattering~\cite{Kawamura2026_2}. Through the self-consistency condition, this nonequilibrium distribution feeds back on the superconducting order parameter. This feedback can suppress or enhance the condensate and may even stabilize unconventional nonequilibrium superconducting states~\cite{Scalapino1977, Hida1978, Gray1978, Kawamura2024, Kawamura2025, Kawamura2026}.

A particularly striking manifestation of this feedback is the microwave-induced enhancement of the critical current $I_{\rm c}$ in superconducting microbridges, commonly known as the Wyatt--Dayem effect~\cite{Wyatt1966, Dayem1967, Klapwijk2020, Tinkham1996}. Subsequent experiments on superconducting films reported microwave-induced enhancement of both the superconducting gap $\Delta$ and the critical temperature $T_{\rm c}$~\cite{Klapwijk1976, Klapwijk1977, Kommers1977, Pals1979, Hall1980}. These observations have motivated broader efforts to understand how nonequilibrium driving, together with relaxation and dissipation, can favor Cooper pairing despite competing heating and pair-breaking processes. Related effects have been investigated in a variety of settings, including $d$-wave superconductors~\cite{Steineman2026}, multiband superconductors~\cite{Isoyama2021, Sobolev2023}, flat-band systems~\cite{Arora2025}, electron-phonon systems~\cite{Knap2016, Sentef2016, Babadi2017}, cavity-coupled superconductors~\cite{Curtis2019, Islam2025}, and ultracold atomic gases~\cite{Robertson2009}.

In conventional superconductors, microwave-induced enhancement is commonly attributed to Eliashberg's kinetic mechanism, in which microwave absorption redistributes quasiparticles away from the gap edge and enhances superconductivity~\cite{Eliashberg1970, Eliashberg1971}. Modern microscopic treatments of this effect are based on the Keldysh--Usadel theory for diffusive superconductors~\cite{Tikhonov2018}. Although this framework successfully describes superconducting enhancement within a limited region of the temperature--drive-frequency plane near $T_{\rm c}$, its validity is restricted to the diffusive limit and thus cannot capture the evolution of microwave-stimulated superconductivity across different elastic impurity-scattering regimes.

For an equilibrium $s$-wave superconductor, Anderson's theorem states that nonmagnetic impurities do not change the superconducting gap $\Delta$ or the critical temperature $T_{\rm c}$~\cite{Anderson1959}. Under microwave irradiation, however, nonmagnetic impurities play a key role. In a conventional single-band BCS superconductor, a spatially uniform microwave field cannot induce single-photon bulk absorption in the absence of momentum relaxation~\cite{Ahn2021,Crowley2022}. Nonmagnetic impurities provide the required momentum relaxation, enabling microwave absorption and reshaping the quasiparticle distribution~\cite{Mattis1958,Steineman2026}. Because the resulting quasiparticle distribution feeds back on the order parameter, nonmagnetic impurities can affect not only $\Delta$ and $T_{\rm c}$ but also the character of the superconducting phase transition.

In the diffusive regime under microwave irradiation, the order parameter $\Delta(T)$ is known to become multivalued near $T_{\rm c}$, leading to bistability and a discontinuous phase transition between the normal and stimulated superconducting states~\cite{Schmid1977,Tikhonov2018}. By contrast, when elastic impurity scattering vanishes, $\Delta(T)$ is expected to remain single-valued because the impurity-assisted absorption channel responsible for the bistability is absent. How the nonequilibrium phase transition changes between these two limits remains unclear. Many early experiments on microwave-stimulated superconductivity were performed on diffusive Al films~\cite{Klapwijk1976,Kommers1977}. Recent advances in epitaxial growth, however, have produced high-quality Al films with substantially reduced and tunable elastic impurity scattering~\cite{Yeh2023,Do2026}. These developments enable experimental studies of the nonequilibrium phase transition beyond the strongly diffusive limit. A unified description across these scattering regimes is thus needed to understand the role of impurities in microwave-stimulated superconductivity.

In this work, we address this problem within a self-consistent Floquet--Nambu--Keldysh formulation. We consider a BCS superconductor coupled to a noninteracting fermionic reservoir that phenomenologically accounts for inelastic relaxation. Such reservoir-coupled setups have been widely used in theoretical studies of driven superconductivity~\cite{Tikhonov2018,Collado2019,Collado2020,Yang2021,Lin2024,Zhang2025,Steineman2026}. Elastic scattering by nonmagnetic impurities is treated within the self-consistent Born approximation. This framework provides a unified description of the driven superconducting phase transition across impurity-scattering regimes, from weak scattering to the diffusive regime.

\begin{figure}[t]
\centering
\includegraphics[width=\linewidth]{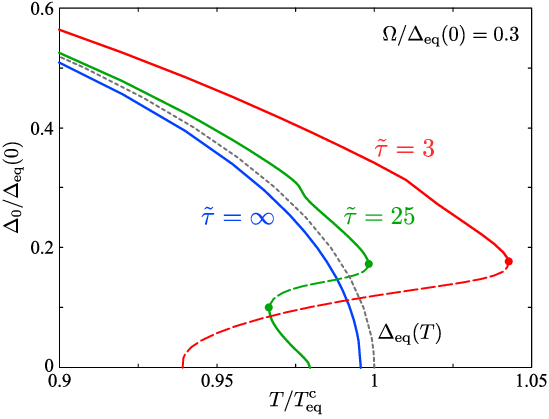}
\caption{
Temperature dependence of the time-averaged superconducting order parameter $\Delta_0(T)$ for several values of the impurity-scattering time $\tilde{\tau}$. The solid and dashed curves represent stable and unstable solutions, respectively, while the dotted curve shows the equilibrium order parameter $\Delta_{\rm eq}(T)$. In the absence of impurity scattering ($\tilde{\tau}=\infty$), $\Delta_0(T)$ is single-valued. At intermediate impurity scattering ($\tilde{\tau}=25$), two additional folds produce a nested multivalued structure with bistability between two superconducting states. At stronger impurity scattering ($\tilde{\tau}=3$), the low-gap stable branch disappears, leaving bistability between the superconducting and normal states. The drive frequency and amplitude are fixed at $\Omega/\Delta_{\rm eq}(0)=0.3$ and $eAv_{\rm F}/\Delta_{\rm eq}(0)=0.2$, respectively. The reservoir-induced damping rate, defined in Eq.~\eqref{eq:gamma_definition}, is fixed at $\gamma/\Delta_{\rm eq}(0)=0.01$.
}
\label{Fig_main_branch} 
\end{figure}

Figure~\ref{Fig_main_branch} illustrates our central finding by showing the temperature dependence of the time-averaged superconducting order parameter $\Delta_0$ at fixed drive amplitude $A$ and frequency $\Omega$ for several values of the dimensionless scattering time $\tilde{\tau}=\tau\Delta_{\rm eq}(0)$. Here, $\Delta_{\rm eq}(T)$ and $T_{\rm eq}^{\rm c}$ denote, respectively, the equilibrium order parameter and critical temperature in the absence of microwave driving. By Anderson's theorem, both $\Delta_{\rm eq}(T)$ and $T_{\rm eq}^{\rm c}$ are independent of the elastic scattering time $\tau$.

In the absence of impurity scattering ($\tilde{\tau}=\infty$), $\Delta_0(T)$ remains single-valued, as expected. At intermediate impurity scattering ($\tilde{\tau}=25$), $\Delta_0(T)$ exhibits a nested multivalued structure, with multiple stable superconducting solutions over a finite temperature range. As impurity scattering is increased further ($\tilde{\tau}=3$), this nested structure evolves into the conventional bistable bifurcation reported in previous studies~\cite{Schmid1977,Tikhonov2018}. Thus, increasing impurity scattering does not merely enlarge the bistable region but qualitatively changes the bifurcation structure itself.

The dotted curve in Fig.~\ref{Fig_main_branch} shows the temperature dependence of the equilibrium order parameter $\Delta_{\rm eq}(T)$ in the absence of microwave driving. For finite $\tilde{\tau}$, the order parameter $\Delta_0$ on the upper superconducting branch exceeds its equilibrium value. At stronger impurity scattering, superconducting solutions also persist above the equilibrium critical temperature $T_{\rm eq}^{\rm c}$. This enhancement originates from the Eliashberg-type redistribution of quasiparticles induced by impurity-assisted microwave absorption. In the absence of impurity scattering ($\tilde{\tau}=\infty$), by contrast, no such enhancement occurs; instead, $\Delta_0$ is slightly suppressed over the entire temperature range. This suppression results from a model-specific absorption process associated with coupling to the fermionic reservoir.

This paper is organized as follows. Section~\ref{sec.formalism} introduces the model and the Floquet--Nambu--Keldysh formalism for driven superconductors. In Secs.~\ref{sec.diffusive} and \ref{sec.inter}, we present the nonequilibrium phase diagrams in the strong- and intermediate-scattering regimes, respectively. Section~\ref{sec.clean} discusses reservoir-assisted microwave absorption in the absence of impurity scattering. Throughout this paper, we set $\hbar=k_{\rm B}=1$.

\section{Formalism}
\label{sec.formalism}

\subsection{Model Hamiltonian}

We consider a spatially uniform, $s$-wave BCS superconductor driven by a time-periodic electromagnetic field. The total Hamiltonian is given by
\begin{equation}
H(t)=H_{\rm SC}(t)+H_{\rm imp}+H_{\rm res}+H_{\rm T}.
\label{eq:Htotal}
\end{equation}
The driven superconductor is described by the mean-field Hamiltonian
\begin{equation}
\scalebox{0.95}{$\displaystyle
H_{\rm SC}(t)
=
\sum_{\bm{k},\sigma}
\xi_{\bm{k}-e\bm A(t)}
a^\dagger_{\bm{k},\sigma}a_{\bm{k},\sigma}
-
\sum_{\bm{k}}
\left[
\Delta(t)
a^\dagger_{\bm{k},\up}
a^\dagger_{-\bm{k},\down}
+{\rm H.c.}
\right].$}
\label{eq:Hsc}
\end{equation}
Here, $a_{\bm{k},\sigma}$ annihilates an electron with momentum $\bm{k}$ and spin $\sigma=\up, \down$, and $\xi_{\bm{k}}=\ep_{\bm{k}}-\mu$ is the single-particle energy measured from the chemical potential. We assume a three-dimensional electron gas with a parabolic dispersion,
\begin{equation}
\ep_{\bm{k}}=\frac{\bm{k}^2}{2m},
\end{equation}
where $m$ is the electron mass.

In Eq.~\eqref{eq:Hsc}, the electromagnetic field is introduced through minimal coupling. We take the vector potential to be
\begin{equation}
\bm{A}(t)
=
A\sin(\Omega t)\,\bm{e}_{\rm p},
\label{eq:vector_potential}
\end{equation}
where $A$ and $\Omega$ are the amplitude and frequency of the drive, respectively, and $\bm{e}_{\rm p}$ is a unit vector along its polarization direction. Under the periodic drive, the superconducting order parameter is determined self-consistently through
\begin{equation}
\Delta(t)
=
U\sum_{\bm{k}}
\braket{
a_{-\bm{k},\downarrow}(t)
a_{\bm{k},\uparrow}(t)
},
\label{eq:gap_definition}
\end{equation}
where $U>0$ is the strength of the attractive pairing interaction.

In Eq.~\eqref{eq:Htotal}, $H_{\rm imp}$ describes elastic scattering by nonmagnetic impurities in the superconductor and is given by
\begin{equation}
H_{\rm imp}
=
u_{\rm imp}
\sum_{\bm{k},\bm{q},\sigma}
\sum_{j=1}^{N_{\rm imp}}
e^{-i \bm{q}\cdot\bm{R}_j}
a^\dagger_{\bm{k}+\bm{q},\sigma}
a_{\bm{k},\sigma}.
\label{eq:Himp}
\end{equation}
Here, $u_{\rm imp}$ is the impurity scattering amplitude, $N_{\rm imp}$ is the number of impurities, and $\bm{R}_j$ denotes the position of the $j$th impurity. We assume that the impurity positions are independently and uniformly distributed.

Under periodic driving, elastic impurity scattering redistributes momentum but cannot dissipate the energy absorbed from the drive. In real superconductors, the absorbed energy is relaxed through inelastic processes such as electron-phonon and electron-electron scattering~\cite{Tinkham1996}. Rather than treating these microscopic relaxation processes explicitly, we model their net effect phenomenologically by coupling the superconductor to an equilibrium fermionic reservoir. Such a reservoir model is widely used to describe relaxation in a driven superconductor~\cite{Tikhonov2018, Collado2019, Collado2020, Yang2021, Lin2024, Zhang2025, Steineman2026}. The reservoir and tunneling Hamiltonians in Eq.~\eqref{eq:Htotal} are given by
\begin{align}
H_{\rm res}
&=
\sum_{\bm{p},\sigma}
\xi^{\rm res}_{\bm{p}}
c^\dagger_{\bm{p},\sigma}
c_{\bm{p},\sigma},
\label{eq:Hres}
\\
H_{\rm T}
&=
\sum_{\bm{p},\bm{k},\sigma}
\left[
\Gamma\,
c^\dagger_{\bm{p},\sigma}
a_{\bm{k},\sigma}
+
{\rm H.c.}
\right].
\label{eq:HT}
\end{align}
Here, $c_{\bm{p},\sigma}$ annihilates a reservoir electron and $\Gamma$ is the tunneling matrix element between the superconductor and the reservoir. We assume that the reservoir is sufficiently large to remain in thermal equilibrium at temperature $T$ and at the same electrochemical potential as the superconductor.

Within this model, we seek a time-periodic steady state in which the time-averaged power supplied by the drive is balanced by the power dissipated into the reservoir. The driven superconductor generally has a nonthermal distribution and cannot be characterized by a unique temperature. Throughout this work, we instead use the reservoir temperature $T$ as the temperature variable, including in the definition of the superconducting critical temperature.

\subsection{Floquet--Nambu--Keldysh Green's function method}
\label{sec.formalism.NEGF}

We introduce the $2\times2$ Nambu Green's functions of the superconductor as
\begin{subequations}
\begin{align}
\hat{G}^{\rm R}(\bm{k},t,t')
&=
-i\Theta(t-t')
\braket{
\big[
\Psi_{\bm{k}}(t),
\Psi^\dagger_{\bm{k}}(t')
\big]_+
},
\\
\hat{G}^{\rm A}(\bm{k},t,t')
&=
\Big[
\hat{G}^{\rm R}(\bm{k},t',t)
\Big]^\dagger,
\\
\hat{G}^<(\bm{k},t,t')
&=
i\braket{
\Psi^\dagger_{\bm{k}}(t')
\Psi_{\bm{k}}(t)
}.
\end{align}
\end{subequations}
Here, $\Theta(t)$ is the Heaviside step function, $[A,B]_+=AB+BA$, and
\begin{equation}
\Psi_{\bm{k}}
=
\begin{pmatrix}
a_{\bm{k},\up}\\[4pt]
a^\dagger_{-\bm{k},\down}
\end{pmatrix}
\end{equation}
is the two-component Nambu field. From the definition of the lesser Green's function and Eq.~\eqref{eq:gap_definition}, the self-consistency condition for the superconducting order parameter can be written as
\begin{equation}
\Delta(t)
=
-iU
\sum_{\bm{k}}
{\rm Tr}
\left[
\hat{\tau}_-
\hat{G}^<(\bm{k},t,t)
\right],
\label{eq:self_consistency}
\end{equation}
where $\hat{\tau}_-=[\hat{\tau}_1-i\hat{\tau}_2]/2$, with $\hat{\tau}_{i=1,2,3}$ being the Pauli matrices in Nambu space.

The Nambu Green's functions obey the Dyson--Keldysh equations~\cite{Rammer2007, HaugJauho2008, Stefanucci2013}:
\begin{align}
&
\hat{G}^{\rm R(A)}(\bm{k},t,t')=
\hat{\m{G}}^{\rm R(A)}(\bm{k},t,t')
+
\int_{-\infty}^{\infty} dt_1
\int_{-\infty}^{\infty} dt_2
\notag\\
&\qquad\times
\hat{\m{G}}^{\rm R(A)}(\bm{k},t,t_1)\,
\hat{\Sigma}^{\rm R(A)}(t_1,t_2)\,
\hat{G}^{\rm R(A)}(\bm{k},t_2,t'),
\label{eq:Dyson}
\\[4pt]
&
\hat{G}^<(\bm{k},t,t')=
\int_{-\infty}^{\infty} dt_1
\int_{-\infty}^{\infty} dt_2\,
\notag\\
&\qquad\times
\hat{G}^{\rm R}(\bm{k},t,t_1)\,
\hat{\Sigma}^{<}(t_1,t_2)\,
\hat{G}^{\rm A}(\bm{k},t_2,t').
\label{eq:Keldysh}
\end{align}
Here, $\hat{\m{G}}^{\rm R(A)}$ is the bare Green's function of the driven superconductor in the absence of impurity scattering and coupling to the reservoir. Its retarded component satisfies~\cite{Rammer2007, HaugJauho2008, Stefanucci2013}
\begin{equation}
\left[
i\frac{\partial}{\partial t}\hat{\tau}_0
-
\hat{\m H}_{\rm SC}(\bm{k},t)
\right]
\hat{\m{G}}^{\rm R}(\bm{k},t,t')
=
\delta(t-t')\, \hat{\tau}_0,
\label{eq:G0R}
\end{equation}
where $\hat{\tau}_0$ is the unit matrix in Nambu space, and
\begin{equation}
\hat{\m H}_{\rm SC}(\bm{k},t)
=
\begin{pmatrix}
\xi_{\bm{k}-e\bm A(t)} & -\Delta(t) \\[4pt]
-\Delta^*(t) & -\xi_{\bm{k}+e\bm A(t)}
\end{pmatrix},
\label{eq.BdG}
\end{equation}
is the Bogoliubov--de Gennes (BdG) Hamiltonian.

The total self-energy contains contributions from the reservoir and impurity scattering,
\begin{equation}
\hat{\Sigma}^{\rm X}(t,t')
=
\hat{\Sigma}^{\rm X}_{\rm res}(t,t')
+
\hat{\Sigma}^{\rm X}_{\rm imp}(t,t'),
\label{eq:self}
\end{equation}
where ${\rm X}={\rm R},{\rm A},<$ denotes the retarded, advanced, and lesser components, respectively. The reservoir self-energy $\hat{\Sigma}^{\rm X}_{\rm res}$ accounts for the phenomenological relaxation induced by the coupling between the superconductor and the reservoir. In the wide-band limit, it is given by~\cite{Kawamura2024,Kawamura2025,Kawamura2026}
\begin{subequations}
\begin{align}
\hat{\Sigma}^{\rm R}_{\rm res}(t,t')
&=
-i\gamma \hat{\tau}_0\, \delta(t-t'),
\label{eq:selfR_res}
\\[4pt]
\hat{\Sigma}^{\rm A}_{\rm res}(t,t')
&=
\big[
\hat{\Sigma}^{\rm R}_{\rm res}(t',t)
\big]^\dagger,
\label{eq:selfA_res}
\\[4pt]
\hat{\Sigma}^<_{\rm res}(t,t')
&=
2i\gamma
\int_{-\infty}^{\infty}
\frac{d\omega}{2\pi}\,
e^{-i\omega(t-t')}
f(\omega)\, \hat{\tau}_0.
\label{eq:selfL_res}
\end{align}
\end{subequations}
Here,
\begin{equation}
f(\omega)
=
\frac{1}{e^{\omega/T}+1}
\end{equation}
is the Fermi--Dirac distribution function in the reservoir, and
\begin{equation}
\gamma
=
\pi N_{\rm res}(0)|\Gamma|^2,
\label{eq:gamma_definition}
\end{equation}
where $N_{\rm res}(0)$ is the density of states of the reservoir at the Fermi level. The parameter $\gamma$ measures the strength of the system--reservoir coupling and sets the phenomenological relaxation scale.

In Eq.~\eqref{eq:self}, $\hat{\Sigma}^{\rm X}_{\rm imp}$ accounts for elastic scattering by nonmagnetic impurities. Within the self-consistent Born approximation, it is given by~\cite{Kawamura2026}
\begin{align}
\hat{\Sigma}^{\rm X}_{\rm imp}(t,t')
&=
n_{\rm imp}u_{\rm imp}^2
\sum_{\bm{k}}
\hat{\tau}_3
\hat{G}^{\rm X}(\bm{k},t,t')
\hat{\tau}_3
\nonumber\\
&=
\frac{1}{2\pi\tau}
\int_{-\infty}^{\infty}d\xi\,
\left\langle
\hat{\tau}_3
\hat{G}^{\rm X}(\xi,\hat{\bm{k}}_{\rm F},t,t')
\hat{\tau}_3
\right\rangle_{\hat{\bm{k}}_{\rm F}},
\label{eq:self_imp}
\end{align}
where $n_{\rm imp}=N_{\rm imp}/V$ is the impurity density. The elastic scattering time $\tau$ is defined by
\begin{equation}
\frac{1}{\tau}
=
2\pi n_{\rm imp}u_{\rm imp}^2N(0),
\end{equation}
where $N(0)$ is the normal-state density of states at the Fermi level. The second line of Eq.~\eqref{eq:self_imp} follows from the constant-density-of-states approximation near the Fermi surface, under which the momentum $\bm{k}$ summation is decomposed into an integral over $\xi$ and an angular average over the Fermi surface,
\begin{equation}
\left\langle F(\hat{\bm{k}}_{\rm F}) \right\rangle_{\hat{\bm{k}}_{\rm F}}
\equiv
\int\frac{d\Omega_{\hat{\bm{k}}_{\rm F}}}{4\pi}\,
F(\hat{\bm{k}}_{\rm F}).
\end{equation}
Here, $\bm{k}_{\rm F}=k_{\rm F}\hat{\bm{k}}_{\rm F}$ is the Fermi momentum, and $\hat{\bm{k}}_{\rm F}$ denotes its direction.

The nonequilibrium order parameter $\Delta(t)$ is obtained by solving the gap, Dyson, and Keldysh equations, Eqs.~\eqref{eq:self_consistency}--\eqref{eq:Keldysh}, self-consistently. Solving these equations directly in the time domain is computationally demanding. We thus use the periodicity of the driven steady state. Specifically, we consider a time-periodic steady state synchronized with the drive, in which the Green's functions and self-energies satisfy
\begin{subequations}
\begin{align}
&
\hat{G}^{\rm X}(\bm{k},t+\m{T},t'+\m{T})=
\hat{G}^{\rm X}(\bm{k},t,t'),
\\[4pt]
&
\hat{\Sigma}^{\rm X}(t+\m{T},t'+\m{T})=
\hat{\Sigma}^{\rm X}(t,t'),
\end{align}
\end{subequations}
where $\m{T}=2\pi/\Omega$ is the driving period. Through the self-consistency condition, this periodicity also implies $\Delta(t+\m{T})=\Delta(t)$. The Dyson and Keldysh equations can then be expressed in Floquet space as~\cite{Zhang2025, Tsuji2008}
\begin{align}
&
\hat{G}^{\rm R(A)}_{m,n}(\bm{k},\omega)
=
\hat{\m{G}}^{\rm R(A)}_{m,n}(\bm{k},\omega)
\notag\\[4pt]
&\qquad
+
\sum_{p,q}
\hat{\m{G}}^{\rm R(A)}_{m,p}(\bm{k},\omega)
\hat{\Sigma}^{\rm R(A)}_{p,q}(\omega)
\hat{G}^{\rm R(A)}_{q,n}(\bm{k},\omega),
\label{eq:Dyson_FL}
\\
&
\hat{G}^{<}_{m,n}(\bm{k},\omega)
=
\sum_{p,q}
\hat{G}^{\rm R}_{m,p}(\bm{k},\omega)
\hat{\Sigma}^{<}_{p,q}(\omega)
\hat{G}^{\rm A}_{q,n}(\bm{k},\omega).
\label{eq:Keldysh_FL}
\end{align}
For a two-time quantity $\hat{F}(t,t')$ satisfying $\hat{F}(t+\m{T},t'+\m{T})=\hat{F}(t,t')$, we define its Floquet representation by
\begin{align}
\hat{F}_{m,n}(\omega)
&=
\frac{1}{\m{T}}
\int_0^{\m{T}}dt_{\rm ave}\,
e^{i(m-n)\Omega t_{\rm ave}}
\notag\\
&\qquad \times
\int_{-\infty}^{\infty}dt_{\rm rel}\,
e^{i[\omega +(m+n)\Omega/2]t_{\rm rel}}
\hat{F}(t,t'),
\end{align}
where
\begin{equation}
t_{\rm ave}=\frac{t+t'}{2},
\qquad
t_{\rm rel}=t-t'.
\end{equation}
Here, $m,n\in\mathbb{Z}$ are the Floquet indices. In this representation, the time convolutions in the Dyson--Keldysh equations become matrix products in Floquet space, substantially reducing the computational cost of the self-consistent calculation.

In the Floquet representation, the reservoir and impurity self-energies in Eqs.~\eqref{eq:selfR_res}-\eqref{eq:selfL_res} and \eqref{eq:self_imp} take the form
\begin{subequations}
\begin{align}
\hat{\Sigma}^{\rm R(A)}_{{\rm res},m,n}(\omega)
&=
\mp i\gamma \hat{\tau}_0\, \delta_{m,n},
\\[4pt]
\hat{\Sigma}^<_{{\rm res},m,n}(\omega)
&=
2i\gamma f(\omega+m\Omega)\hat{\tau}_0\, \delta_{m,n},
\\[4pt]
\hat{\Sigma}^{\rm X}_{{\rm imp},m,n}(\omega)
&=
n_{\rm imp}u_{\rm imp}^2
\sum_{\bm{k}}
\hat{\tau}_3
\hat{G}^{\rm X}_{m,n}(\bm{k},\omega)
\hat{\tau}_3.
\end{align}
\end{subequations}
We note that the reservoir self-energy is diagonal in Floquet space because the reservoir itself is undriven, whereas the impurity self-energy generally acquires off-diagonal Floquet components through the Green's function of the driven superconductor.

In Floquet space, the equation for the bare retarded Green's function, Eq.~\eqref{eq:G0R}, becomes
\begin{equation}
\sum_l
\left[
[\omega+m\Omega]\hat{\tau}_0\, \delta_{m,l}
-
\hat{\m H}_{{\rm SC},m,l}(\bm{k})
\right]
\hat{\m{G}}^{\rm R}_{l,n}(\bm{k},\omega)
=
\delta_{m,n}\hat{\tau}_0.
\end{equation}
Here, the Floquet representation of the BdG Hamiltonian $\hat{\m H}_{\rm SC}(\bm{k},t)$ defined in Eq.~\eqref{eq.BdG} is given by
\begin{align}
\hat{\m H}_{{\rm SC},m,n}(\bm{k})
&=
\frac{1}{\m{T}}
\int_{-\m{T}/2}^{\m{T}/2}dt\,
e^{i(m-n)\Omega t}\,
\hat{\m H}_{\rm SC}(\bm{k},t)
\notag\\
&\equiv
\begin{pmatrix}
\xi^-_{\bm{k},m-n} & -\Delta_{m-n} \\[4pt]
-\Delta^*_{n-m} & -\xi^+_{\bm{k},m-n}
\end{pmatrix}.
\label{eq.HSC.mn}
\end{align}
The quantity $\Delta_j$ denotes the $j$th Floquet component of the order parameter,
\begin{equation}
\Delta_j
=
\frac{1}{\m{T}}
\int_{-\m{T}/2}^{\m{T}/2}dt\,
e^{ij\Omega t}\Delta(t).
\end{equation}
Its zeroth component,
\begin{equation}
\Delta_0
=
\frac{1}{\m{T}}
\int_{-\m{T}/2}^{\m{T}/2}dt\,\Delta(t),
\label{eq.def.D0}
\end{equation}
is the time-averaged superconducting order parameter.

In Eq.~\eqref{eq.HSC.mn}, $\xi^\pm_{\bm{k},j}$ denote the $j$th Floquet components of $\xi_{\bm{k}\pm e\bm{A}(t)}$ and are defined by
\begin{equation}
\xi^\pm_{\bm{k},j}
=
\frac{1}{\m T}
\int_{-\m T/2}^{\m T/2}dt\,
e^{ij\Omega t}\,
\xi_{\bm{k}\pm e\bm A(t)}.
\label{eq:xi_Floquet_definition}
\end{equation}
Near the Fermi surface, we expand the field-dependent dispersion to first order in the vector potential, which gives
\begin{equation}
\xi_{\bm{k} \pm e\bm{A}(t)}
\simeq
\xi_{\bm{k}}
\pm
eAv_F\,
\hat{\bm{k}}_{\rm F} \cdot
\bm{e}_{\rm p}\,
\sin(\Omega t).
\label{eq:xi_continuum_time}
\end{equation}
Here, $v_{\rm F}$ is the Fermi velocity. This approximation is valid when the drive-induced energy scale $eAv_F$ is much smaller than the Fermi energy $\ep_F$, as is the case throughout the parameter range considered in this work. The corresponding Floquet components are then given by
\begin{equation}
\xi^\pm_{\bm{k},j}
=
\xi_{\bm{k}}\, \delta_{j,0}
\pm
\frac{ieAv_F}{2}\,
\hat{\bm{k}}_{\rm F} \cdot\bm{e}_{\rm p}\,
\big[\delta_{j,1}-\delta_{j,-1}\big].
\label{eq:xi_Floquet_continuum}
\end{equation}

The Floquet components of the order parameter follow from the self-consistency condition. Substituting the Floquet representation into Eq.~\eqref{eq:self_consistency}, we obtain~\cite{Zhang2025}
\begin{align}
\Delta_j
&=
-iU
\sum_{\bm{k}}
\sum_l
\int_{-\Omega/2}^{\Omega/2}
\frac{d\omega}{2\pi}\,
{\rm Tr}
\left[
\hat{\tau}_-
\hat{G}^<_{l+j,l}(\bm{k},\omega)
\right]
\notag\\
&=
-iU
\sum_{\bm{k}}
\int_{-\infty}^{\infty}
\frac{d\omega}{2\pi}\,
{\rm Tr}
\left[
\hat{\tau}_-
\hat{G}^<_{j,0}(\bm{k},\omega)
\right].
\label{eq:self_consistent_FL}
\end{align}
To obtain the second line, we use the redundancy relation of the Floquet representation~\cite{Tsuji2008}
\begin{equation}
\hat{G}^<_{m+l,n+l}(\omega-l\Omega)
=
\hat{G}^<_{m,n}(\omega),
\end{equation}
which follows directly from the definition of the Floquet components. This identity combines the sum over the Floquet index $l$ and the frequency integral over the first Floquet Brillouin zone into a single extended-zone integral over the entire real axis.

We solve the Floquet Dyson and Keldysh equations, Eqs.~\eqref{eq:Dyson_FL} and \eqref{eq:Keldysh_FL}, together with the self-consistency equation~\eqref{eq:self_consistent_FL} to determine the Green's functions, the impurity self-energy, and the Floquet components $\Delta_j$ of the order parameter self-consistently. In the numerical calculations, the Floquet indices are truncated to $-N_{\rm F}\le m,n\le N_{\rm F}$, with $N_{\rm F}$ chosen sufficiently large to ensure convergence with respect to the Floquet truncation. Details of the numerical procedure are provided in Appendix~\ref{app:numerical_calculations}.

We note that the self-consistency equation~\eqref{eq:self_consistent_FL} may yield multiple steady-state solutions for the same set of parameters. In equilibrium, their thermodynamic stability can be assessed from the free-energy landscape. Such an approach is generally unavailable out of equilibrium, where stability should, strictly speaking, be determined by explicitly following the real-time evolution~\cite{Kawamura2024,Kawamura2025} or by performing a linear stability analysis around each steady-state solution~\cite{Hanai2017,Kawamura2022}. Here, we instead adopt a simpler operational criterion that has been used in previous stusy of nonequilibrium superconductivity~\cite{Ouassou2018}. Starting from a slightly perturbed configuration near each solution, we iterate the self-consistency equation using simple mixing. If the iteration converges back to the original solution, we classify it as stable under this criterion. If the deviation grows under iteration, we classify the solution as unstable. For specific models, this criterion has been shown to yield stability classifications consistent with those obtained from explicit real-time evolution~\cite{Snyman2009, Ouassou2018}.

\section{Eliashberg-type enhancement induced by impurity scattering}
\label{sec.diffusive}

\begin{figure}[t]
\centering
\includegraphics[width=7.8cm]{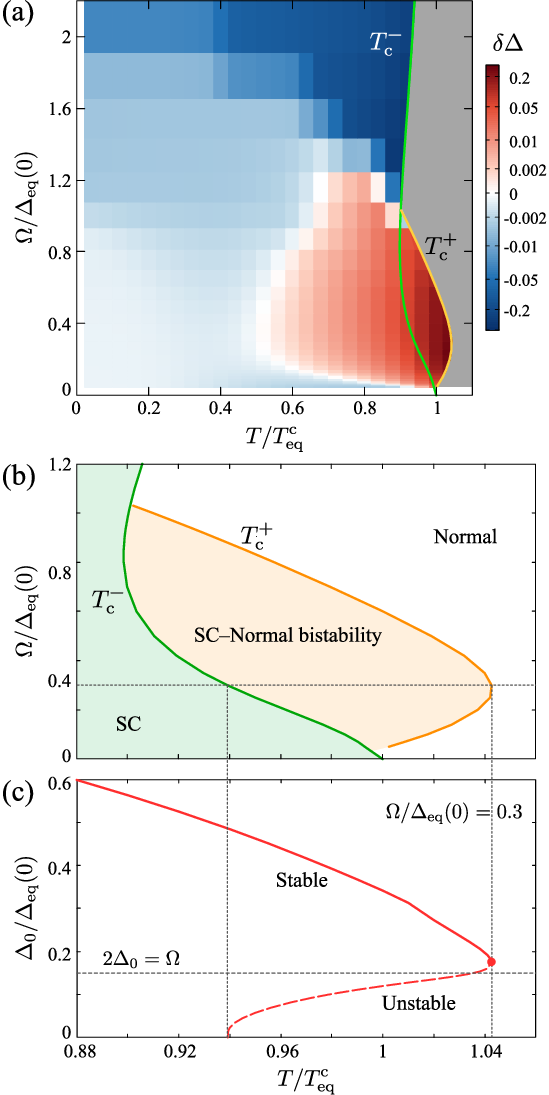}
\caption{
Microwave-induced enhancement and nonequilibrium phase structure in the strong-scattering regime ($\tilde{\tau}=3$). (a) Microwave-induced change in the time-averaged superconducting order parameter, as quantified by $\delta\Delta$ defined in Eq.~\eqref{eq:delta_Delta}, in the $T$--$\Omega$ plane. The red and blue regions correspond to enhancement ($\delta\Delta>0$) and suppression ($\delta\Delta<0$), respectively. The gray-shaded region denotes the normal phase, where no superconducting solution is obtained. (b) Nonequilibrium phase diagram near $T_{\rm eq}^{\rm c}$. (c) Temperature dependence of the time-averaged order parameter $\Delta_0(T)$ at $\Omega/\Delta_{\rm eq}(0)=0.3$. The horizontal dotted line marks the single-photon pair-breaking threshold, $2\Delta_0=\Omega$.
}
\label{Fig_tau3_phase} 
\end{figure}

We first consider the strong-scattering regime ($\tilde{\tau}=3$), where impurity-assisted microwave absorption produces a pronounced enhancement of the superconducting order parameter. In this regime, our approach recovers the main qualitative features of microwave-stimulated superconductivity reported in Ref.~\cite{Tikhonov2018}. In that work, the microwave field is treated perturbatively within the Keldysh--Usadel formalism, and the frequency--temperature phase diagram is determined in the weak-power limit. By contrast, our Floquet formulation treats the periodic drive nonperturbatively, allowing us to determine the superconducting response at finite microwave powers beyond the leading-order expansion. In addition, Ref.~\cite{Tikhonov2018} assumes a quasi-one-dimensional diffusive wire, whereas our calculation retains the angular dependence over the three-dimensional Fermi surface. Despite these differences, the two approaches yield qualitatively consistent results in the strong-scattering regime.

Unless otherwise stated, all results presented below are obtained at fixed vector-potential amplitude $eAv_{\rm F}/\Delta_{\rm eq}(0)=0.2$ and reservoir coupling strength $\gamma/\Delta_{\rm eq}(0)=0.01$. Here, $\Delta_{\rm eq}(T)$ denotes the equilibrium superconducting order parameter at temperature $T$ in the absence of the drive and, by Anderson's theorem, is independent of the impurity-scattering time $\tilde{\tau}$.

Figure~\ref{Fig_tau3_phase}(a) shows the microwave-induced change in the time-averaged superconducting order parameter in the $T$--$\Omega$ plane. The normalized change is defined as
\begin{equation}
\delta\Delta
=
\frac{\Delta_0-\Delta_{\rm eq}(T)}
{\Delta_{\rm eq}(0)},
\label{eq:delta_Delta}
\end{equation}
where $\Delta_0$ is the time-averaged superconducting order parameter under microwave irradiation, as defined in Eq.~\eqref{eq.def.D0}. In the red region ($\delta\Delta>0$), the microwave drive enhances superconductivity, whereas in the blue region ($\delta\Delta<0$), it suppresses superconductivity. The microwave-induced enhancement is confined to a bounded region near the equilibrium critical temperature $T_{\rm eq}^{\rm c}$, in qualitative agreement with the phase diagram obtained in previous work~\cite{Tikhonov2018, Derendorf2024}.

\begin{figure}[t]
\centering
\includegraphics[width=7.5cm]{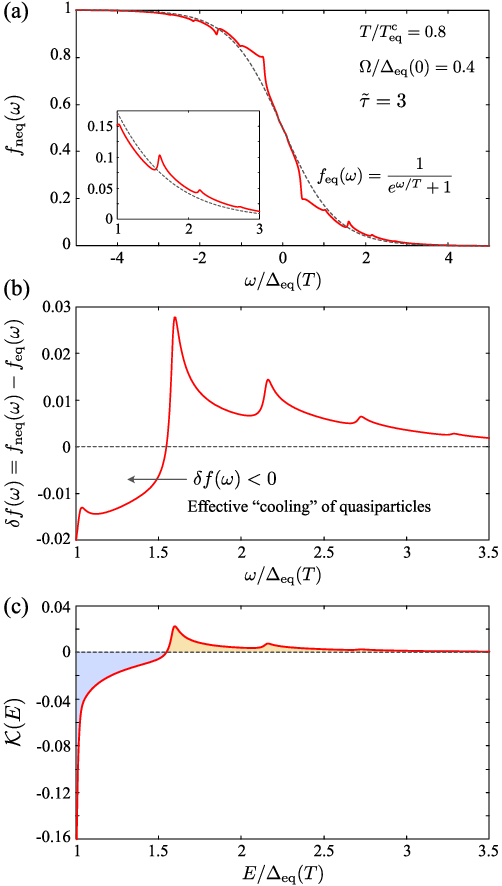}
\caption{
Nonequilibrium quasiparticle redistribution underlying the microwave-induced enhancement. (a) Time-averaged nonequilibrium distribution function $f_{\rm neq}(\omega)$ compared with the equilibrium Fermi--Dirac distribution $f_{\rm eq}(\omega)$. (b) Drive-induced change $\delta f(\omega)=f_{\rm neq}(\omega)-f_{\rm eq}(\omega)$ above the equilibrium gap edge. (c) Weighted distribution change $\mathcal{K}(E)$ defined in Eq.~\eqref{eq:pairing_kernel}. The region with $\mathcal{K}(E)<0$ near the gap edge (blue shaded area) gives a positive contribution to the order parameter, whereas the region with $\mathcal{K}(E)>0$ at higher energies (orange shaded area) gives a negative contribution. We set $T/T_{\rm eq}^{\rm c}=0.8$ and $\Omega/\Delta_{\rm eq}(0)=0.4$.
}
\label{Fig_tau3_fneq} 
\end{figure}

To clarify the microscopic origin of the enhancement near $T_{\rm eq}^{\rm c}$, Fig.~\ref{Fig_tau3_fneq}(a) shows a representative time-averaged nonequilibrium distribution function $f_{\rm neq}(\omega)$ in the enhancement region. Within the Floquet convention introduced in Sec.~\ref{sec.formalism.NEGF}, this distribution function is extracted from the lesser Green's function as
\begin{equation}
f_{\rm neq}(\omega)
=
\frac{-i}{2\pi N_{\rm neq}(\omega)}
\sum_{\bm{k}}
{\rm Tr}
\left[
\hat{\tau}_{\rm p}
\hat{G}^{<}_{0,0}(\bm{k},\omega)
\right],
\label{eq:fneq}
\end{equation}
where $\hat{\tau}_{\rm p}=[\hat{\tau}_0+\hat{\tau}_3]/2$ projects onto the particle sector. The corresponding time-averaged single-particle density of states is obtained from the retarded component as
\begin{equation}
N_{\rm neq}(\omega)
=
-\frac{1}{\pi}
{\rm Im}
\sum_{\bm{k}}
{\rm Tr}
\left[
\hat{\tau}_{\rm p}
\hat{G}^{\rm R}_{0,0}(\bm{k},\omega)
\right].
\label{eq:dos_time_averaged}
\end{equation}

As shown in Fig.~\ref{Fig_tau3_fneq}(a), the nonequilibrium distribution function $f_{\rm neq}(\omega)$ deviates from the Fermi--Dirac distribution of the undriven system,
\begin{equation}
f_{\rm eq}(\omega)
=
\frac{1}{1+e^{\omega/T}}.
\end{equation}
Whereas $f_{\rm eq}(\omega)$ decreases monotonically with energy $\omega$, $f_{\rm neq}(\omega)$ exhibits a nonmonotonic structure. A similar nonmonotonic quasiparticle distribution under microwave irradiation was reported in Ref.~\cite{Visser2014}. Figure~\ref{Fig_tau3_fneq}(b) shows the drive-induced change
\begin{equation}
\delta f(\omega)
=
f_{\rm neq}(\omega)-f_{\rm eq}(\omega)
\end{equation}
for energies above the undriven gap edge, $\omega>\Delta_{\rm eq}(T)$. We find that $\delta f(\omega)<0$ immediately above the gap, whereas $\delta f(\omega)>0$ at higher energies.  This behavior can be interpreted as an effective ``cooling" of quasiparticles near the gap edge, resulting from the microwave-induced redistribution of quasiparticles from energies just above the gap to higher energies.

To understand how this cooling near the gap edge affects the superconducting order parameter, it is useful to recall the equilibrium BCS gap equation~\cite{Eliashberg1970},
\begin{equation}
\Delta
=
UN(0)
\int_{\Delta}^{\omega_{\rm D}}dE\,
\frac{\Delta}{\sqrt{E^2-\Delta^2}}
\left[1-2f_{\rm eq}(E)\right].
\label{eq:BCS_principal}
\end{equation}
To isolate the effect of the nonequilibrium distribution, we retain the equilibrium spectral kernel and linearize Eq.~\eqref{eq:BCS_principal} around $\Delta_{\rm eq}(T)$. Within this approximation, the kinetic contribution to the change in the gap can be written as
\begin{equation}
\Delta_0-\Delta_{\rm eq}(T)
\simeq
-\mathcal{C}(T)
\int_{\Delta_{\rm eq}(T)}^{\omega_{\rm D}}dE\,
\frac{\delta f(E)}
{\sqrt{E^2-\Delta_{\rm eq}^2(T)}},
\label{eq:delta_Delta_kinetic}
\end{equation}
where $\mathcal{C}(T)>0$ is a response coefficient arising from the linearization of the equilibrium gap equation around $\Delta_{\rm eq}(T)$. Figure~\ref{Fig_tau3_fneq}(c) shows the integrand
\begin{equation}
\mathcal{K}(E)
=
\frac{\delta f(E)}
{\sqrt{E^2-\Delta_{\rm eq}^2(T)}}.
\label{eq:pairing_kernel}
\end{equation}
The blue region with $\mathcal{K}(E)<0$ immediately above the gap gives a positive contribution to $\Delta_0-\Delta_{\rm eq}(T)$, whereas the orange region with $\mathcal{K}(E)>0$ at higher energies suppresses the gap. Because the BCS kernel is strongly weighted toward the gap edge, the depletion of quasiparticles near the gap dominates over their accumulation at higher energies. This provide a microscopic demonstration of the kinetic mechanism proposed by Eliashberg, in which superconductivity is enhanced through the nonequilibrium redistribution of thermally excited quasiparticles.~\cite{Eliashberg1970, Eliashberg1971}.

The quasiparticle redistribution responsible for the gap enhancement also modifies the superconducting phase transition near $T_{\rm eq}^{\rm c}$. Figure~\ref{Fig_tau3_phase}(b) shows the nonequilibrium phase diagram in the $T$--$\Omega$ plane, while Fig.~\ref{Fig_tau3_phase}(c) shows the corresponding bifurcation diagram of $\Delta_0(T)$ at $\Omega/\Delta_{\rm eq}(0)=0.3$. Below the lower boundary $T_{\rm c}^{-}$, the superconducting state ($\Delta_0>0$) is the only stable state, whereas above the upper boundary $T_{\rm c}^{+}$, only the normal state ($\Delta_0=0$) is stable. Between these boundaries, both states are stable, resulting in a bistable region previously found in the diffusive limit~\cite{Schmid1977,Tikhonov2018}.

The upper boundary $T_{\rm c}^{+}$ exceeds the undriven critical temperature $T_{\rm eq}^{\rm c}$ over a finite frequency range. The stimulated superconducting state therefore persists above the equilibrium transition temperature, indicating an enhancement of the superconducting critical temperature. Upon heating, the superconducting state disappears discontinuously at $T_{\rm c}^{+}$, whereas upon cooling, the normal state loses stability at $T_{\rm c}^{-}$. This bifurcation gives rise to a first-order-like phase transition with hysteresis under temperature cycling, consistent with early experimental observations~\cite{Klapwijk1977,Pals1979}.

The bifurcation in Fig.~\ref{Fig_tau3_phase}(c) is closely related to the onset of single-photon pair breaking. In the presence of impurity scattering, a photon can break a Cooper pair and create two quasiparticles when the photon energy exceeds the pair-breaking threshold, $\Omega>2\Delta_0$. The horizontal dashed line in Fig.~\ref{Fig_tau3_phase}(c) marks this threshold condition $2\Delta_0=\Omega$. Since the resulting quasiparticle excitation strongly suppresses superconductivity, the stable superconducting branch remains in the region $2\Delta_0>\Omega$ and terminates before entering the pair-breaking regime.

The same energy scale approximately determines the upper-frequency boundary of the enhancement region in Fig.~\ref{Fig_tau3_phase}(a). Below the pair-breaking threshold, microwave absorption redistributes thermally excited quasiparticles and enhances the order parameter. Once the single-photon pair-breaking channel opens, quasiparticle generation dominates and suppresses superconductivity. Thus, the region with $\delta\Delta>0$ is approximately bounded by the threshold $\Omega=2\Delta_{\rm eq}(T)$.

\begin{figure}[t]
\centering
\includegraphics[width=\linewidth]{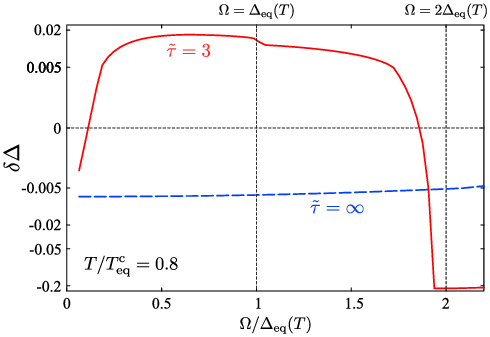}
\caption{
Dependence of $\delta\Delta$ on the drive frequency $\Omega$ at $T/T_{\rm eq}^{\rm c}=0.8$. The vertical axis uses a symmetric logarithmic scale. The solid and dashed curves correspond to $\tilde{\tau}=3$ and $\tilde{\tau}=\infty$, respectively. For $\tilde{\tau}=3$, the weak dip around $\Omega=\Delta_{\rm eq}(T)$ and the strong suppression near $\Omega=2\Delta_{\rm eq}(T)$ are associated with two-photon and single-photon pair breaking, respectively.
}
\label{Fig_T08_Omega_scan} 
\end{figure}

To see this frequency dependence more closely, Fig.~\ref{Fig_T08_Omega_scan} shows $\delta\Delta$ as a function of the drive frequency $\Omega$ at $T/T_{\rm eq}^{\rm c}=0.8$. For $\tilde{\tau}=3$, the gap is enhanced over a broad frequency range below the pair-breaking threshold. As $\Omega$ approaches $2\Delta_{\rm eq}(T)$, $\delta\Delta$ decreases sharply and becomes strongly negative due to quasiparticle generation by single-photon pair breaking.

A weaker dip is also visible around $\Omega=\Delta_{\rm eq}(T)$. At this frequency, the combined energy of two photons reaches the pair-breaking threshold, $2\Omega=2\Delta_{\rm eq}(T)$. This condition corresponds to a two-photon process in which a Cooper pair absorbs two photons and breaks into two quasiparticles. We attribute the weak dip around $\Omega=\Delta_{\rm eq}(T)$ to this two-photon pair-breaking process. Its much smaller magnitude compared with the strong suppression near $\Omega=2\Delta_{\rm eq}(T)$ is consistent with the higher-order nature of the two-photon process.

We note that, in the absence of impurity scattering ($\tilde{\tau}=\infty$), the dashed curve in Fig.~\ref{Fig_T08_Omega_scan} remains smooth near both $\Omega=\Delta_{\rm eq}(T)$ and $\Omega=2\Delta_{\rm eq}(T)$, in contrast to the result for $\tilde{\tau}=3$. This behavior reflects the absence of impurity-assisted microwave absorption processes that produce the corresponding pair-breaking features. The remaining weak and smooth suppression originates from the coupling to the fermionic reservoir, as discussed in Sec.~\ref{sec.clean}.

\section{Nested bifurcation at intermediate impurity scattering}
\label{sec.inter}

We now turn to the intermediate-scattering regime, where impurity-assisted microwave absorption produces a qualitatively different bifurcation structure. As shown in Sec.~\ref{sec.diffusive}, microwave absorption can redistribute thermally excited quasiparticles and enhance superconductivity, but it can also induce pair breaking and suppress superconductivity. The self-consistent feedback of these competing processes on the order parameter $\Delta_0(T)$ gives rise to a rich nonequilibrium phase structure. As summarized in Fig.~\ref{Fig_main_branch}, increasing the impurity-scattering strength first transforms the single-valued $\Delta_0(T)$ into a nested multivalued structure, which subsequently evolves into the conventional bistability characteristic of the diffusive limit. In this section, we analyze this intermediate regime using $\tilde{\tau}=25$ as a representative case.

\begin{figure}[t]
\centering
\includegraphics[width=7.5cm]{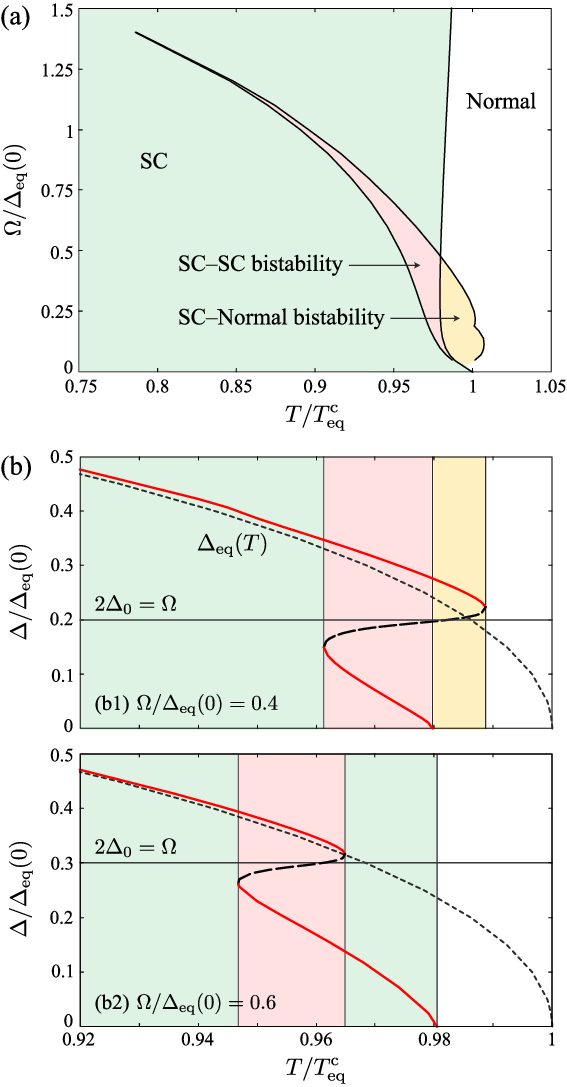}
\caption{
Nonequilibrium phase diagram and bifurcation structure in the intermediate-scattering regime ($\tilde{\tau}=25$). (a) Phase diagram in the $T$--$\Omega$ plane. The yellow-shaded region corresponds to bistability between the superconducting and normal states, while the red-shaded region corresponds to bistability between two superconducting states with different values of $\Delta_0$. (b1), (b2) Bifurcation diagrams of $\Delta_0(T)$ at $\Omega/\Delta_{\rm eq}(0)=0.4$ and $0.6$, respectively. The solid red curves represent stable superconducting solutions, while the black dashed curves represent unstable solutions. The dotted curves show the equilibrium order parameter $\Delta_{\rm eq}(T)$, and the horizontal lines mark the single-photon pair-breaking threshold, $2\Delta_0=\Omega$.
}
\label{Fig_inter_phase} 
\end{figure}

Figure~\ref{Fig_inter_phase}(a) shows the phase diagram in the $T$--$\Omega$ plane for $\tilde{\tau}=25$. In addition to the regions in which only the superconducting or normal state is stable, two distinct types of bistability appear. The yellow-shaded region at relatively low frequencies corresponds to the conventional bistability between a superconducting state and the normal state discussed in Sec.~\ref{sec.diffusive}. At higher frequencies, the red-shaded region corresponds to bistability between two stable superconducting states with different values of $\Delta_0$.

The origin of these two bistable regions can be understood from the bifurcation diagram in Fig.~\ref{Fig_inter_phase}(b1), shown for $\Omega/\Delta_{\rm eq}(0)=0.4$. The solid red curves represent stable superconducting solutions, while the black dashed curve represents an unstable solution. The dotted curve shows the equilibrium order parameter $\Delta_{\rm eq}(T)$ in the absence of the drive. The high-gap branch lies above the equilibrium curve, $\Delta_0>\Delta_{\rm eq}(T)$, whereas the low-gap branch lies below it, $\Delta_0<\Delta_{\rm eq}(T)$. Thus, the two branches correspond to microwave-enhanced and microwave-suppressed superconducting states, respectively.

At low temperatures ($T/T_{\rm eq}^{\rm c}\lesssim0.96$), only the high-gap superconducting state is stable. As the temperature is increased, a stable low-gap branch and an unstable branch emerge at the first fold, giving rise to the red-shaded region in which both superconducting states are stable. At a higher temperature ($T/T_{\rm eq}^{\rm c}\gtrsim0.98$), the low-gap branch reaches the normal state at $\Delta_0=0$, and the system enters the yellow-shaded region in which both the high-gap superconducting state and the normal state are stable. Finally, the high-gap branch merges with the unstable branch at the second fold and disappears, leaving only the normal state.

A different sequence appears at the higher frequency $\Omega/\Delta_{\rm eq}(0)=0.6$, as shown in Fig.~\ref{Fig_inter_phase}(b2). Starting from the high-gap superconducting state at low temperatures, a stable low-gap branch and an unstable branch again emerge at the first fold, producing a finite temperature range with two stable superconducting solutions. At the second fold, however, the high-gap branch merges with the unstable branch and disappears, leaving the low-gap branch as the only stable superconducting solution. This branch then decreases continuously to $\Delta_0=0$. Thus, along this frequency cut, the bistability between the two superconducting states lies entirely within the superconducting phase and is not accompanied by bistability between the superconducting and normal states.

These bifurcation diagrams also indicate the phase transitions expected under a quasistatic temperature sweep. For $\Omega/\Delta_{\rm eq}(0)=0.4$, the high-gap superconducting state persists upon heating until it disappears at the upper fold, where the system switches discontinuously to the normal state. Upon cooling, the normal state first undergoes a continuous transition to the low-gap superconducting state. This state remains stable down to the lower fold, where the system switches discontinuously to the high-gap superconducting state. The resulting sequence exhibits both normal--superconducting hysteresis and an additional switching between the two superconducting states.

For $\Omega/\Delta_{\rm eq}(0)=0.6$, the normal--superconducting transition remains continuous upon both heating and cooling. Within the superconducting phase, however, the high-gap state switches discontinuously to the low-gap state at the upper fold upon heating, whereas the low-gap state switches back to the high-gap state at the lower fold upon cooling. The red-shaded region thus produces a distinct hysteresis between two superconducting states entirely within the superconducting phase.

The horizontal lines in Fig.~\ref{Fig_inter_phase}(b) mark the single-photon pair-breaking threshold, $2\Delta_0=\Omega$, which separates the two stable superconducting branches. While the single-photon pair-breaking channel is energetically closed on the high-gap branch ($2\Delta_0>\Omega$), thermally excited quasiparticles above the gap can absorb photons and thereby be redistributed away from the gap edge. This redistribution contributes to the Eliashberg-type enhancement discussed in Sec.~\ref{sec.diffusive}. On the low-gap branch ($2\Delta_0<\Omega$), by contrast, the photon energy exceeds the pair-breaking threshold. Single-photon absorption can then break Cooper pairs and generate additional quasiparticles, leading to a quasiparticle-rich state with a strongly suppressed order parameter.

This contrast between the two branches gives rise to a positive feedback mechanism. At a fixed drive frequency $\Omega$, a decrease in $\Delta_0$ such that $2\Delta_0<\Omega$ activates pair breaking, increases the quasiparticle population, and further suppresses the gap. Conversely, an increase in $\Delta_0$ such that $2\Delta_0>\Omega$ closes the pair-breaking channel and allows quasiparticle redistribution to dominate, leading to further enhancement through the Eliashberg-type kinetic mechanism. When this feedback is sufficiently strong, the intermediate solution near the threshold becomes unstable, while stable high- and low-gap solutions remain on either side. As a result, the solution curve of the self-consistency equation develops two folds, producing the bistability between the two superconducting states shown in Fig.~\ref{Fig_inter_phase}(b).

The evolution with increasing impurity-scattering strength, summarized in Fig.~\ref{Fig_main_branch}, can also be understood from this feedback picture. When impurity scattering is weak, impurity-assisted absorption is insufficient to generate strong feedback around the pair-breaking threshold, and $\Delta_0(T)$ remains single-valued. At intermediate scattering, the feedback becomes sufficiently strong to produce two additional folds and the nested multivalued structure. Increasing the scattering rate further strengthens both quasiparticle redistribution and pair breaking, with different consequences for the two stable branches. On the high-gap branch, the pair-breaking channel remains closed, and the stronger redistribution of thermally excited quasiparticles increases $\Delta_0$. On the low-gap branch, by contrast, the open pair-breaking channel generates more quasiparticles and further suppresses $\Delta_0$ until the low-gap state can no longer be sustained as a stable solution. The bistability between the two superconducting states then disappears, leaving the conventional bistability between the high-gap superconducting state and the normal state.

\section{Reservoir-assisted microwave absorption without impurity scattering}
\label{sec.clean}

In the preceding sections, we showed that impurity-assisted microwave absorption can enhance superconductivity through quasiparticle redistribution, induce pair breaking, and lead to bistability. We now consider the case without impurity scattering ($\tilde{\tau}=\infty$). In a spatially uniform continuum superconductor, the single-photon absorption channel mediated by impurity scattering is absent~\cite{Ahn2021,Crowley2022}. However, as shown in Figs.~\ref{Fig_main_branch} and \ref{Fig_T08_Omega_scan}, the microwave drive suppresses the superconducting gap below its equilibrium value even at $\tilde{\tau}=\infty$. We explain below that this residual suppression originates from a separate absorption process enabled by coupling to the fermionic reservoir used to describe inelastic relaxation.

\begin{figure}[t]
\centering
\includegraphics[width=\linewidth]{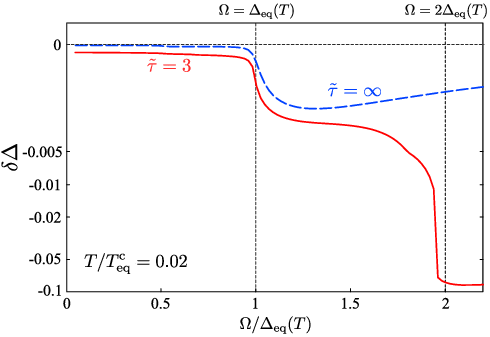}
\caption{
Dependence of $\delta\Delta$ on the drive frequency $\Omega$ at $T/T_{\rm eq}^{\rm c}=0.02$. The vertical axis uses a symmetric logarithmic scale, as in Fig.~\ref{Fig_T08_Omega_scan}. The solid and dashed curves correspond to $\tilde{\tau}=3$ and $\tilde{\tau}=\infty$, respectively. The pronounced structure near $\Omega=\Delta_{\rm eq}(T)$ persists in the absence of impurity scattering, whereas the structure near $\Omega=2\Delta_{\rm eq}(T)$ appears only at finite impurity scattering.
}
\label{Fig_T002_Omega_scan} 
\end{figure}

To clarify the origin of this residual suppression, we examine its frequency dependence at a low temperature, $T/T_{\rm eq}^{\rm c}=0.02$. Figure~\ref{Fig_T002_Omega_scan} compares $\delta\Delta(\Omega)$ for $\tilde{\tau}=\infty$ and $\tilde{\tau}=3$. For $\tilde{\tau}=\infty$, a pronounced suppression appears around $\Omega=\Delta_{\rm eq}(T)$, whereas no corresponding structure is found near the ordinary single-photon pair-breaking threshold, $\Omega=2\Delta_{\rm eq}(T)$. For $\tilde{\tau}=3$, by contrast, structures appear around both frequencies. The feature at $\Omega=2\Delta_{\rm eq}(T)$ originates from impurity-assisted single-photon pair breaking, while the feature at $\Omega=\Delta_{\rm eq}(T)$ persists even without impurity scattering, indicating a different absorption mechanism.

This feature can be understood as a consequence of single-particle exchange between the superconductor and the fermionic reservoir. Because the reservoir can supply or remove individual fermions, microwave absorption does not necessarily create two quasiparticles within the superconductor. Instead, the combined action of the drive and the reservoir coupling can excite a fermion near the Fermi level of the reservoir ($\omega=0$) into a quasiparticle state above the superconducting gap. The minimum photon energy required for this process is approximately $\Delta_{\rm eq}(T)$ rather than $2\Delta_{\rm eq}(T)$, giving rise to the structure around $\Omega=\Delta_{\rm eq}(T)$. This reservoir-assisted absorption remains possible even at $\tilde{\tau}=\infty$ and suppresses the order parameter by increasing the quasiparticle population.

The reservoir-assisted feature is strongly temperature dependent. At low temperatures, the sharp Fermi edge of the reservoir produces a well-defined absorption threshold: a fermion occupying a state near the reservoir Fermi level can be excited into the superconducting quasiparticle continuum once $\Omega$ reaches approximately $\Delta_{\rm eq}(T)$. When the temperature becomes comparable to the superconducting gap scale, $T\sim\Delta_{\rm eq}(T)$, thermal broadening of the reservoir distribution smears the distinction between occupied initial states and unoccupied final states. As a result, the reservoir-assisted absorption no longer exhibits a well-defined threshold at $\Omega\simeq\Delta_{\rm eq}(T)$. This explains why the feature is pronounced in Fig.~\ref{Fig_T002_Omega_scan} but is not clearly resolved at $T/T_{\rm eq}^{\rm c}=0.8$ in Fig.~\ref{Fig_T08_Omega_scan}.

This temperature dependence is also important for distinguishing reservoir-assisted absorption from two-photon pair breaking. At low temperatures, the structure around $\Omega=\Delta_{\rm eq}(T)$ appears both with and without impurity scattering, showing that the position of the feature alone is insufficient to identify it as a two-photon process. At $T/T_{\rm eq}^{\rm c}=0.8$, by contrast, the $\tilde{\tau}=\infty$ curve varies smoothly around this frequency, whereas a dip remains visible for $\tilde{\tau}=3$. We attribute this additional impurity-dependent feature to two-photon pair breaking, for which the combined energy of two absorbed photons satisfies the threshold condition $2\Omega=2\Delta_{\rm eq}(T)$. Thus, the structure near $\Omega=\Delta_{\rm eq}(T)$ can contain contributions from different absorption processes, depending on temperature and impurity scattering.

The reservoir-assisted absorption identified here is not merely an artifact of the fermionic-reservoir model. Single-particle exchange can occur when a superconductor is coupled through an interface to a normal-metal reservoir, a quasiparticle trap, or another electronic continuum~\cite{Riwar2016,Silveri2017}. Indeed, photon-assisted single-electron tunneling has been observed in normal-metal--insulator--superconductor junctions~\cite{Tan2017}. In such systems, the fermionic-reservoir model can capture a genuine photon-assisted excitation process involving particle exchange between the superconductor and its environment.

For a bulk superconductor, however, energy relaxation to the environment is typically mediated by electron--phonon scattering~\cite{Kaplan1976}. A phonon bath relaxes the electronic distribution without directly supplying or removing electrons and thus does not generically produce the single-particle absorption channel with the threshold $\Omega\simeq\Delta_{\rm eq}(T)$. The low-temperature response obtained using a fermionic-reservoir model should therefore be interpreted with caution when applied to the intrinsic microwave response of a bulk superconductor.

Finally, we emphasize that reservoir-assisted absorption is not responsible for the nested bifurcation discussed in Sec.~\ref{sec.inter}. The reservoir-assisted feature persists in the absence of impurity scattering and is most pronounced around $\Omega\simeq\Delta_{\rm eq}(T)$ at low temperatures. By contrast, the nested bifurcation occurs only at intermediate impurity scattering near $T_{\rm eq}^{\rm c}$. Its high- and low-gap branches are separated by the single-photon pair-breaking threshold $2\Delta_0=\Omega$, rather than by the reservoir-assisted condition $\Delta_0=\Omega$. The reservoir-assisted process thus does not alter the physical interpretation of our main result. Instead, it highlights a separate limitation of fermionic-reservoir models, which are widely used in theories of driven superconductivity~\cite{Tikhonov2018,Collado2019,Collado2020,Yang2021,Lin2024,Zhang2025,Steineman2026}.

\section{Summary}

In summary, we have investigated the nonequilibrium steady states of a periodically driven BCS superconductor in the presence of nonmagnetic impurity scattering and energy relaxation. Using the Floquet--Nambu--Keldysh Green's function method together with the self-consistent Born approximation, we have self-consistently determined the superconducting order parameter under microwave driving.

We found that the nonequilibrium phase diagram depends strongly on the impurity-scattering strength. When impurity scattering is absent or weak, the steady-state order parameter is single-valued at each temperature and drive frequency. At intermediate impurity scattering, bistability emerges between microwave-enhanced and microwave-suppressed superconducting states. At stronger impurity scattering, this evolves into bistability between the superconducting and normal states, as found in the diffusive limit~\cite{Schmid1977,Tikhonov2018}. This evolution reflects competition between two effects of impurity-assisted microwave absorption: Eliashberg-type enhancement due to quasiparticle redistribution and suppression through pair-breaking-induced quasiparticle generation

We also identified a reservoir-assisted absorption channel arising from single-particle exchange with the fermionic reservoir. This channel generates excess quasiparticles and suppresses the superconducting order parameter even in the absence of impurity scattering. However, it is distinct from the mechanism responsible for the impurity-dependent evolution of the bifurcation structure and phase diagram. Our main result thus cannot be attributed to this model-specific absorption channel. Rather, it highlights a separate limitation of fermionic-reservoir models for driven bulk superconductors.

An important direction for future work is to treat electron--phonon relaxation explicitly. Unlike a fermionic reservoir, a phonon bath relaxes the quasiparticle distribution without exchanging particles with the superconductor and thus does not introduce the reservoir-assisted absorption channel. Such a treatment would also capture the strong energy dependence of quasiparticle relaxation and recombination times~\cite{Kaplan1976}. In particular, the slow relaxation of quasiparticles near the gap edge may make the drive-induced quasiparticle redistribution more pronounced and further strengthen the Eliashberg-type enhancement. Extending the present framework to include microscopic electron--phonon relaxation would thus enable a more quantitative description of driven bulk superconductors.

\begin{acknowledgments}
We thank Yusuke Kato for stimulating discussions. T.K. was supported by MEXT and JSPS KAKENHI Grant-in-Aid for JSPS fellows Grant No.~25K23363. Some of the computations in this work were done using the facilities of the Supercomputer Center, the Institute for Solid State Physics, the University of Tokyo.
\end{acknowledgments}

\section*{Data Availability}
The data that support the findings of this article are not publicly available. The data are available from the authors upon reasonable request.

\appendix
\section{Details of the numerical calculations}
\label{app:numerical_calculations}

This appendix describes the numerical procedure used to solve the Floquet Dyson--Keldysh equations~\eqref{eq:Dyson_FL} and \eqref{eq:Keldysh_FL} together with the self-consistency equation~\eqref{eq:self_consistent_FL}. We use a check accent to denote matrices acting in the combined Floquet--Nambu space. Each block of such a matrix, denoted by a hat, is a $2\times2$ matrix in Nambu space. In this representation, the Green's functions are expressed as
\begin{equation}
\check{G}^{\rm X}(\bm{k},\omega)
=
\begin{pmatrix}
\ddots & \vdots & \vdots & \vdots & \iddots \\
\cdots &
\hat{G}^{\rm X}_{-1,-1} &
\hat{G}^{\rm X}_{-1,0} &
\hat{G}^{\rm X}_{-1,1} &
\cdots \\[4pt]
\cdots &
\hat{G}^{\rm X}_{0,-1} &
\hat{G}^{\rm X}_{0,0} &
\hat{G}^{\rm X}_{0,1} &
\cdots \\[4pt]
\cdots &
\hat{G}^{\rm X}_{1,-1} &
\hat{G}^{\rm X}_{1,0} &
\hat{G}^{\rm X}_{1,1} &
\cdots \\
\iddots & \vdots & \vdots & \vdots & \ddots
\end{pmatrix}.
\end{equation}
In this notation, the Floquet Dyson--Keldysh equations~\eqref{eq:Dyson_FL} and \eqref{eq:Keldysh_FL} take the compact matrix form
\begin{align}
&
\check{G}^{\rm R(A)}(\bm{k},\omega)=
\check{\m{G}}^{\rm R(A)}(\bm{k},\omega)
\notag\\
&\qquad+
\check{\m{G}}^{\rm R(A)}(\bm{k},\omega)\,
\check{\Sigma}^{\rm R(A)}(\omega)\,
\check{G}^{\rm R(A)}(\bm{k},\omega),
\label{eq:Dyson.app1}
\\[4pt]
&
\check{G}^{<}(\bm{k},\omega)=
\check{G}^{\rm R}(\bm{k},\omega)\,
\check{\Sigma}^{<}(\omega)\,
\check{G}^{\rm A}(\bm{k},\omega).
\label{eq:Keldysh.app1}
\end{align}

\subsection{$\xi$-integrated Green's functions}

We define the $\xi$-integrated Green's functions by
\begin{equation}
\check{\mathfrak{G}}^{\rm X}
(\hat{\bm{k}}_{\rm F},\omega)
\equiv
\int_{-\infty}^{\infty}d\xi\,
\check{G}^{\rm X}
(\xi,\hat{\bm{k}}_{\rm F},\omega).
\label{eq:xi_integrated_GF}
\end{equation}
Because the self-energy is independent of $\xi$, the Keldysh equation~\eqref{eq:Keldysh.app1} gives
\begin{align}
\check{\mathfrak{G}}^{<}(\hat{\bm{k}}_{\rm F},\omega)
&=
\int_{-\infty}^{\infty}d\xi\,
\check{G}^{\rm R}(\xi,\hat{\bm{k}}_{\rm F},\omega)\,
\check{\Sigma}^{<}(\omega)\,
\check{G}^{\rm A}(\xi,\hat{\bm{k}}_{\rm F},\omega)
\notag\\[4pt]
&\equiv
\m{I}\!\left[\check{\Sigma}^{<}(\omega)\right].
\label{eq:integrated_lesser_bilinear}
\end{align}

The $\xi$-integrated Green's functions $\check{\mathfrak{G}}^{\rm X}$ enter both the self-consistency equation and the impurity self-energy. In terms of $\check{\mathfrak{G}}^<$, Eq.~\eqref{eq:self_consistent_FL} becomes
\begin{equation}
\Delta_j
=
-iUN(0)
\int_{-\infty}^{\infty}
\frac{d\omega}{2\pi}
\Braket{
{\rm Tr}
\left[
\hat{\tau}_{-}
\hat{\mathfrak{G}}^{<}_{j,0}
(\hat{\bm{k}}_{\rm F},\omega)
\right]
}_{\hat{\bm{k}}_{\rm F}}.
\label{eq:self_consistent_xi_integrated}
\end{equation}
The impurity self-energy follows similarly from Eq.~\eqref{eq:self_imp}:
\begin{equation}
\hat{\Sigma}^{\rm X}_{{\rm imp},m,n}(\omega)
=
\frac{1}{2\pi\tau}
\Braket{
\hat{\tau}_3
\hat{\mathfrak{G}}^{\rm X}_{m,n}
(\hat{\bm{k}}_{\rm F},\omega)
\hat{\tau}_3
}_{\hat{\bm{k}}_{\rm F}}.
\label{eq:impurity_self_energy_xi_integrated}
\end{equation}

The self-consistent calculation thus requires an efficient evaluation of the integrated retarded Green's function $\check{\mathfrak{G}}^{\rm R}$, which enters the retarded impurity self-energy, and of the bilinear integral $\m{I}[\check{\Sigma}^{<}(\omega)]$, which yields $\check{\mathfrak{G}}^{<}$. We describe both evaluations below.

\subsection{Resolvent representation of the retarded Green's function}

We first express the retarded Green's function in a form that makes its $\xi$ dependence explicit. The Dyson equation~\eqref{eq:Dyson.app1} gives
\begin{equation}
\check{G}^{\rm R}(\xi,\hat{\bm{k}}_{\rm F},\omega)
=
\left[
\check{M}_{\rm R}(\hat{\bm{k}}_{\rm F},\omega)
-
\xi\check{T}
\right]^{-1},
\label{eq:GR_linear_pencil}
\end{equation}
where
\begin{align}
\big[\check{T}\big]_{m,n}
&=
\hat{\tau}_3\,\delta_{m,n},
\label{eq:T_floquet_nambu}
\\[4pt]
\big[
\check{M}_{\rm R}(\hat{\bm{k}}_{\rm F},\omega)
\big]_{m,n}
&=
[\omega+m\Omega]\hat{\tau}_0\,\delta_{m,n}
\notag\\
&\quad-
\hat{h}_{m,n}(\hat{\bm{k}}_{\rm F})
-
\hat{\Sigma}^{\rm R}_{m,n}(\omega).
\label{eq:MR_floquet_nambu}
\end{align}
Here, $\hat{h}_{m,n}$ is the $\xi$-independent part of the Floquet BdG Hamiltonian. It is obtained by separating out the term $\xi\hat{\tau}_3\delta_{m,n}$ from $\hat{\m H}_{{\rm SC},m,n}$ in Eq.~\eqref{eq.HSC.mn} and is given explicitly by
\begin{align}
\hat{h}_{m,n}(\hat{\bm{k}}_{\rm F})
={}&
-\frac{ieAv_F}{2}
\hat{\bm{k}}_{\rm F}\cdot\bm e_{\rm p}\,
\hat{\tau}_0\,\delta_{m,n+1}
\notag\\
&\quad+
\frac{ieAv_F}{2}
\hat{\bm{k}}_{\rm F}\cdot\bm e_{\rm p}\,
\hat{\tau}_0\,\delta_{m,n-1}
\notag\\
&\quad-
\Delta_{m-n}\hat{\tau}_+
-
\Delta_{n-m}^{*}\hat{\tau}_-.
\label{eq:h_floquet_nambu}
\end{align}

We next rewrite the retarded Green's function in a resolvent form suitable for evaluating $\check{\mathfrak{G}}^{\rm R}$. Using $\check{T}^2=\check{1}$, the matrix in Eq.~\eqref{eq:GR_linear_pencil} can be factorized as
\begin{equation}
\check{M}_{\rm R}(\hat{\bm{k}}_{\rm F},\omega)
-
\xi\check{T}
=
\check{T}
\left[
\check{T}\check{M}_{\rm R}(\hat{\bm{k}}_{\rm F},\omega)
-
\xi\check{1}
\right],
\label{eq:MR_factorization}
\end{equation}
where $\check{1}$ is the identity matrix in the combined Floquet--Nambu space. Inverting this relation and using $\check{T}^{-1}=\check{T}$, we obtain
\begin{equation}
\check{G}^{\rm R}(\xi,\hat{\bm{k}}_{\rm F},\omega)
=
\left[
\check{N}_{\rm R}(\hat{\bm{k}}_{\rm F},\omega)
-
\xi\check{1}
\right]^{-1}
\check{T},
\label{eq:GR_resolvent}
\end{equation}
where
\begin{equation}
\check{N}_{\rm R}(\hat{\bm{k}}_{\rm F},\omega)
\equiv
\check{T}\check{M}_{\rm R}(\hat{\bm{k}}_{\rm F},\omega).
\label{eq:NR_definition}
\end{equation}
This resolvent representation provides the starting point for the efficient evaluation of the $\xi$-integrated Green's functions.

\subsection{Evaluation of the $\xi$ integrals}

We now describe how to evaluate the $\xi$-integrated Green's functions $\check{\mathfrak{G}}^{\rm X}$. In the following, the dependence on $\hat{\bm{k}}_{\rm F}$ and $\omega$ is suppressed for brevity. The evaluation proceeds in four steps.

\begin{enumerate}
\item
We construct an invertible transformation $\check{V}$ that block diagonalizes $\check{N}_{\rm R}$ into invariant subspaces associated with eigenvalues in the upper and lower half-planes of the complex $\xi$ plane:
\begin{equation}
\check{V}^{-1}
\check{N}_{\rm R}
\check{V}
=
\begin{pmatrix}
N_+ & 0\\
0 & N_-
\end{pmatrix}.
\label{eq:NR_invariant_subspaces}
\end{equation}
The eigenvalues of $N_+$ and $N_-$ lie in the upper and lower half-planes, respectively. We denote the identity matrices acting in the corresponding invariant subspaces by $I_+$ and $I_-$, respectively. Numerically, the transformation $\check{V}$ is constructed using an ordered Schur decomposition followed by the solution of a Sylvester equation.

\item
Using the resulting block-resolvent representation, we obtain the $\xi$-integrated retarded and advanced Green's functions as
\begin{align}
\check{\mathfrak{G}}^{\rm R}
&=
-i\pi
\check{V}
\begin{pmatrix}
I_+ & 0\\
0 & -I_-
\end{pmatrix}
\check{V}^{-1}\check{T},
\label{eq:integrated_GR_block}
\\
\check{\mathfrak{G}}^{\rm A}
&=
\left(
\check{\mathfrak{G}}^{\rm R}
\right)^\dagger.
\label{eq:integrated_GA_block}
\end{align}

\item
To evaluate the bilinear integral $\m{I}\big[\check{\Sigma}^<\big]$, we transform the kernel $\check{T}\check{\Sigma}^<\check{T}$ to the same invariant-subspace basis:
\begin{align}
\check{\widetilde{\Sigma}}^<
&=
\check{V}^{-1}
\check{T}\,
\check{\Sigma}^<\,
\check{T}
\big(\check{V}^{-1}\big)^\dagger
\notag\\
&=
\begin{pmatrix}
\widetilde{\Sigma}^<_{++} & \widetilde{\Sigma}^<_{+-}\\[4pt]
\widetilde{\Sigma}^<_{-+} & \widetilde{\Sigma}^<_{--}
\end{pmatrix}.
\label{eq:transformed_lesser_self_energy}
\end{align}
Given $N_\pm$ and $\widetilde{\Sigma}^<_{\pm\pm}$, we determine the two matrices $\m{I}_{++}$ and $\m{I}_{--}$ by solving the Sylvester equations
\begin{equation}
N_\pm\m{I}_{\pm\pm}
-
\m{I}_{\pm\pm}N_\pm^\dagger
=
\pm 2\pi i\,
\widetilde{\Sigma}^<_{\pm\pm}.
\label{eq:Sylvester_Ipm}
\end{equation}
The upper sign applies to $\m{I}_{++}$, whereas the lower sign applies to $\m{I}_{--}$. These Sylvester equations have unique solutions because the spectra of $N_\pm$ and $N_\pm^\dagger$ lie in opposite half-planes.

\item
Finally, we transform the remaining diagonal blocks back to the original Floquet--Nambu basis:
\begin{equation}
\m{I}\big[\check{\Sigma}^<\big]
=
\check{V}
\begin{pmatrix}
\m{I}_{++} & 0\\
0 & \m{I}_{--}
\end{pmatrix}
\check{V}^{\dagger}.
\label{eq:bilinear_xi_integral_result}
\end{equation}
The resulting matrix is the integrated lesser Green's function,
\begin{equation}
\check{\mathfrak{G}}^<
=
\m{I}\!\left[\check{\Sigma}^<\right].
\end{equation}
\end{enumerate}

This procedure reduces the evaluation of the $\xi$-integrated Green's functions to an ordered Schur decomposition and the solution of Sylvester equations. Both operations can be performed efficiently using standard LAPACK routines. The remainder of this subsection describes the individual steps in detail.

We first describe the construction of the transformation $\check{V}$ in Eq.~\eqref{eq:NR_invariant_subspaces}. An ordered complex Schur decomposition of $\check{N}_{\rm R}$ gives
\begin{equation}
\check{Q}^{\dagger}
\check{N}_{\rm R}
\check{Q}
=
\begin{pmatrix}
N_+ & R\\
0 & N_-
\end{pmatrix},
\label{eq:NR_ordered_Schur}
\end{equation}
where $\check{Q}$ is unitary. The eigenvalues of the upper-triangular blocks $N_+$ and $N_-$ lie in the upper and lower half-planes of the complex $\xi$ plane, respectively. The matrix $R$ is the off-diagonal block coupling the two Schur blocks. We compute this decomposition using the LAPACK routine \texttt{ZGEES}.

To eliminate $R$ in Eq.~\eqref{eq:NR_ordered_Schur}, we introduce the block-upper-triangular transformation
\begin{equation}
\check{W}
=
\begin{pmatrix}
I_+ & -Y\\
0 & I_-
\end{pmatrix},
\label{eq:W_definition}
\end{equation}
where $I_\pm$ are the identity matrices in the corresponding Schur subspaces, and $Y$ is an unknown rectangular matrix. Applying this transformation to Eq.~\eqref{eq:NR_ordered_Schur} gives
\begin{equation}
\check{W}^{-1}
\begin{pmatrix}
N_+ & R\\
0 & N_-
\end{pmatrix}
\check{W}
=
\begin{pmatrix}
N_+
&
R-N_+Y+YN_-
\\
0 & N_-
\end{pmatrix}.
\label{eq:Schur_block_transformation}
\end{equation}
Requiring the off-diagonal block on the right-hand side to vanish yields the Sylvester equation
\begin{equation}
N_+Y-YN_-=R.
\label{eq:Sylvester_invariant_subspaces}
\end{equation}
This equation has a unique solution because the spectra of $N_+$ and $N_-$ lie in disjoint half-planes. We solve Eq.~\eqref{eq:Sylvester_invariant_subspaces} for $Y$ using the LAPACK routine \texttt{ZTRSYL}. Finally, we define
\begin{equation}
\check{V}
=
\check{Q}\check{W},
\label{eq:V_definition}
\end{equation}
which gives the block diagonalization in Eq.~\eqref{eq:NR_invariant_subspaces}.

Having constructed $\check{V}$, we substitute Eq.~\eqref{eq:NR_invariant_subspaces} into Eq.~\eqref{eq:GR_resolvent} to obtain
\begin{equation}
\check{G}^{\rm R}(\xi)
=
\check{V}
\begin{pmatrix}
\big[N_+-\xi I_+\big]^{-1} & 0\\
0 & \big[N_--\xi I_-\big]^{-1}
\end{pmatrix}
\check{V}^{-1}\check{T}.
\label{eq:GR_block_resolvent}
\end{equation}
The finite reservoir-induced broadening $\gamma>0$ keeps the spectrum of $\check{N}_{\rm R}$ away from the real axis. Because the spectra of $N_+$ and $N_-$ lie in the upper and lower half-planes, respectively, the resolvent integral gives~\cite{Higham2008}
\begin{equation}
\int_{-\infty}^{\infty}d\xi\,
\big[N_\pm-\xi I_\pm\big]^{-1}
=
\mp i\pi I_\pm.
\label{eq:block_resolvent_integrals}
\end{equation}
Substituting Eq.~\eqref{eq:block_resolvent_integrals} into Eq.~\eqref{eq:GR_block_resolvent} yields Eq.~\eqref{eq:integrated_GR_block}.

We next use the same block-resolvent representation to evaluate $\check{\mathfrak{G}}^<$, which enters the self-consistency equation~\eqref{eq:self_consistent_xi_integrated}. Substituting the block-resolvent forms of the retarded and advanced Green's functions into Eq.~\eqref{eq:integrated_lesser_bilinear}, we obtain
\begin{equation}
\m{I}\big[\check{\Sigma}^<\big]
=
\check{V}
\begin{pmatrix}
\m{I}_{++} & \m{I}_{+-}\\[4pt]
\m{I}_{-+} & \m{I}_{--}
\end{pmatrix}
\check{V}^{\dagger},
\label{eq:bilinear_integral_blocks}
\end{equation}
where
\begin{equation}
\m{I}_{ab}
=
\int_{-\infty}^{\infty}d\xi\,
\big[N_a-\xi I_a\big]^{-1}
\widetilde{\Sigma}^<_{ab}
\big[N_b^\dagger-\xi I_b\big]^{-1}.
\label{eq:Iab_definition}
\end{equation}
Here, $\widetilde{\Sigma}^<_{ab}$ is defined in Eq.~\eqref{eq:transformed_lesser_self_energy}. For $\m{I}_{+-}$, the poles of both resolvents lie in the upper half-plane, whereas for $\m{I}_{-+}$, they lie in the lower half-plane. Because the integrands decay as $1/\xi^2$, the integration contour can be closed in the half-plane containing no poles. The two off-diagonal block integrals thus vanish:
\begin{equation}
\m{I}_{+-}
=
\m{I}_{-+}
=
0.
\label{eq:offdiagonal_integrals_zero}
\end{equation}

It remains to determine the two diagonal blocks, $\m{I}_{++}$ and $\m{I}_{--}$. Contour integration shows that they satisfy the Sylvester equations in Eq.~\eqref{eq:Sylvester_Ipm}. At fixed $\hat{\bm{k}}_{\rm F}$ and $\omega$, and for a given $\check{\Sigma}^<$, the matrices $N_\pm$ and $\widetilde{\Sigma}^<_{\pm\pm}$ are known. Because the spectra of $N_\pm$ and $N_\pm^\dagger$ lie in opposite half-planes, the corresponding Sylvester equations have unique solutions. We determine $\m{I}_{++}$ and $\m{I}_{--}$ numerically using the LAPACK routine \texttt{ZTRSYL}. The bilinear integral, and hence the integrated lesser Green's function $\check{\mathfrak{G}}^<$, is then reconstructed in the original Floquet--Nambu basis using Eq.~\eqref{eq:bilinear_xi_integral_result}.

\subsection{Overall self-consistent procedure}

We finally summarize the overall self-consistent procedure. The calculation starts from initial guesses for the order-parameter components $\Delta_j$ and the impurity self-energies, and proceeds as follows.

\begin{enumerate}
\item
At fixed $\Delta_j$ and $\omega$, and for the current impurity self-energies, we evaluate the $\xi$-integrated retarded and lesser Green's functions for each direction $\hat{\bm{k}}_{\rm F}$ using the procedure described above. The invariant-subspace decomposition used to evaluate $\check{\mathfrak{G}}^{\rm R}$ is reused in the evaluation of $\check{\mathfrak{G}}^<$.

\item
The angular averages of $\check{\mathfrak{G}}^{\rm R}$ and $\check{\mathfrak{G}}^{<}$ are used to update $\check{\Sigma}_{\rm imp}^{\rm R}$ and $\check{\Sigma}_{\rm imp}^{<}$, respectively, through Eq.~\eqref{eq:impurity_self_energy_xi_integrated}. The two impurity self-energies are updated simultaneously until the residual satisfies the convergence criterion.

\item
Once the impurity self-energies have converged at all frequencies, the order-parameter components are updated using Eq.~\eqref{eq:self_consistent_xi_integrated}.

\item
With the updated order parameter, the impurity self-energy is recalculated, and this nested procedure is repeated until the gap equation converges.
\end{enumerate}
Both the inner iteration for the impurity self-energy and the outer iteration for the order parameter are accelerated using Anderson mixing~\cite{Anderson1965}.

\bibliography{Stimulated_SC}
\end{document}